\documentclass[a4paper]{article}
\usepackage[utf8]{inputenc}
\usepackage[T1]{fontenc}
\def\be{\begin{equation}}
\def\ee{\end{equation}}
\newcommand{\g}{\,\mathrm{GeV}}
\newcommand{\m}{M_{H^{\pm}}}
\newcommand{\rg}{R_{\gamma\gamma}}

\newcommand{\lczp}{\lambda_{345}}
\usepackage{amsmath}
\usepackage{amssymb}
\usepackage{amsthm}
\usepackage{amsfonts}
\usepackage{graphicx}
\usepackage[sort]{cite}
\usepackage{authblk}
\begin{document}
\title{Universe in the  light of LHC
\thanks{Presented by M. Krawczyk at the Applied Nuclear Physics and Innovative Technologies 2014 Symposium.}}
\author{M.~Krawczyk, M.~Matej, D.~Soko\l owska, B.~\'{S}wie\.{z}ewska}
\affil{Faculty of Physics, University of Warsaw \\ Pasteura 5, 02-093 Warsaw, Poland}
\date{December 30, 2014}
\maketitle            
\begin{abstract}
The Large Hadron Collider (LHC) provides data  which give information on dark matter.  In particular,  measurements related to the Higgs sector lead to strong  constraints on the invisible sector which are competitive with  astrophysical limits.  Some recent LHC results on  dark matter coming from the Higgs sector in  the Inert Doublet Model (IDM) are  presented. 
\end{abstract}

%
%
\section{Introduction}

In the autumn of the year  2014 one can safely  conclude that the SM-like Higgs  scenario~\cite{BE,Higgs,Kibble}  is being observed at the LHC \cite{ATLAS-Higgs,CMS-Higgs}. Such a scenario  can be realized  in various models beyond the Standard Model (SM). It was found recently that the LHC is very effective in constraining models with so called  Higgs-portal to the dark matter (DM), see eg.~\cite{portalH}. In particular, LHC results on the  Higgs boson properties  can give stronger limits on the Higgs-DM couplings than the astrophysical DM  experiments. In addition, some important constraints are coming from the dedicated search of dark matter at the LHC~\cite{LHCinv}.

Let us start with a little bit of history. It was only 50 years ago when the Quark Model as well as the mass generation mechanism had  been proposed. These were  crucial steps towards building a theory of elementary particles known as the Standard Model.  The first idea of a global SU(3) symmetry arose from an observation and classification of  a plethora of hadrons, which were being  discovered copiously in early 1950s.
It led to our current understanding of the structure of matter at the fundamental level, probed up to  a distance of 10$^{-18}$~cm, in the form of 3 generations of quarks and leptons.

The masses of these particles show no clear pattern, besides the fact that the second  generation of fermions  is heavier than the first one, and the third one is the heaviest.  Among the particles which are carriers of fundamental interactions, the  photon (electromagnetic interaction) and the gluons (strong interaction) are massless, while  $W^{\pm}$ and $Z$, the carriers of the electroweak force,  are massive, actually very massive as compared to the proton. It was already known in the 1960s that this may create a serious problem in describing  a very short range weak interaction (a point-like interaction according to  Fermi) in a theoretical  approach based on a local symmetry.

\subsection{Brout-Englert-Higgs  mechanism in the SM}

The Brout-Englert-Higgs mechanism (B-E-H), proposed already in 1960s, is based on spontaneous breaking of the EW symmetry $SU(2) \times U(1)$ to $U(1)_{\mathrm{QED}}$\cite{BE, Higgs,Kibble}. One SU(2) doublet $\Phi$ of spin 0 fields  with a non-zero vacuum expectation value $v$ (VEV) is introduced, and the gauge bosons and fermions acquire mass thanks to interaction with this constant field component. { Mass of $W^\pm$  generated in this way is equal to  $ M_W = g v/2$, and at tree level $\rho=\frac{M_W}{M_Z \cos \theta_W}  =1$. Masses of fermions are generated due to  Yukawa interaction with  $\Phi$.}

The Higgs boson $h$ which arises in the B-E-H mechanism  has spin 0, even CP parity and no electric charge. Its couplings to the SM particles are all fixed, being  proportional  to their masses. 
The only unknown parameter  is (was)  $M_h$ related to the strength of Higgs  self-interaction. Long term  hunting for  a Higgs boson seems to have reached its finale  in the summer 2012, when at the LHC the Higgs boson with mass around 125~GeV  has been discovered.  Up to now, with all collected data (already 1 million of  Higgses!),  the signal strengths  in various channels (defined with respect to the SM prediction)  are close to 1, and the observed scenario can be described as  a SM-like Higgs scenario.

\subsection{SM-like Higgs scenarios at the LHC}

Although the SM is in  very good agreement with existing data,  there are many serious arguments to go beyond it. The SM has  many free parameters, contains massless  neutrinos, does not have  a DM candidate, cannot describe baryon asymmetry of the Universe, etc. 
The recently discovered 125~GeV scalar 
has properties very close to those predicted by the SM.  But how close?  As long as other new particles are not seen at the LHC the only relevant BSM models  are those allowing for  SM-like scenario, i.e., with
a SM-like Higgs boson and other new particles too heavy or too weakly interacting to be observed in existing experiments. 

The main production channel of the Higgs particle at the LHC is gluon-gluon fusion.
The channels allowing most precise measurements are Higgs decays to $\gamma \gamma$ and $ZZ$.  
Loop couplings of the Higgs to gauge bosons  $gg$, $\gamma \gamma $, $\gamma Z$ are sensitive to new physics (even to contribution of very heavy particles due to nondecoupling effects).
The overall signal strength is equal to $\mu= 1.00 \pm 0.13$ (CMS)~\cite{CMS-Higgs}, $1.30\pm 0.12 \mathrm{(stat)}^{+0.14}_{-0.11}\mathrm{(syst)}$ (ATLAS)~\cite{ATLAS-Higgs}.

\subsection{Dark Matter}

Throughout the years  much evidence for the existence of DM has been collected: rotation curves of galaxies, gravitational lensing, etc.~\cite{dm-corfu}. A typical candidate for DM is the so-called WIMP (weakly interacting massive particle). The DM relic density is inferred from the measurements made by WMAP and Planck with a good accuracy~\cite{dm-relic}. 
There are other astrophysical experiments searching for DM, either directly (via scattering off nuclei) or indirectly (search for products of DM annihilation or decay). Unfortunately, the picture given by these  experiments is not entirely consistent. However, some information about DM can be drawn from the LHC measurements, and hopefully it can shed some light on its nature.

\section{TheInert Doublet Model} 

Among the simplest extensions of the Higgs sector in the SM are models with two  SU(2) doublets (Two Higgs Doublet Models -- 2HDMs). In the non-supersymmetric 2HDMs  a special role is played  by the Inert Doublet Model (IDM) -- the  only version of 2HDM with a stable particle (scalar)
~\cite{Ma, Barbieri}.

In the IDM The scalars' interactions are defined by the following potential
\begin{align}
V=&-\frac{1}{2}\Big[m_{11}^{2}(\phi_{S}^{\dagger}\phi_{S})+m_{22}^{2}(\phi_{D}^{\dagger}\phi_{D})\Big]+\frac{1}{2}\Big[\lambda_{1}(\phi_{S}^{\dagger}\phi_{S})^{2}+\lambda_{2}(\phi_{D}^{\dagger}\phi_{D})^{2}\Big]\\ 
&+\lambda_{3}(\phi_{S}^{\dagger}\phi_{S})(\phi_{D}^{\dagger}\phi_{D})+\lambda_{4}(\phi_{S}^{\dagger}\phi_{D})(\phi_{D}^{\dagger}\phi_{S})+\frac{1}{2}\lambda_{5}\Big[(\phi_{S}^{\dagger}\phi_{D})^{2}+(\phi_{D}^{\dagger}\phi_{S})^{2}\Big].\nonumber
\end{align}
This potential possesses a global discrete $\mathbb{Z}_2$-type  symmetry $D$ under an action of which the field $\phi_D$ changes sign, while $\phi_S$ remains untouched. The interactions with fermions are chosen in the IDM such as to preserve this symmetry,~i.e., only $\phi_S$ couples to fermions. In this way the whole IDM lagrangian  is $D$-symmetric, and moreover,  the vacuum state of this model is such that $D$ is not broken spontaneously. The VEVs of the two doublets read
\begin{equation}
 \langle\phi_S\rangle= \left( \begin{array}{c} 0\\
\frac{v_S}{\sqrt 2}\end{array} \right),\quad
 \langle\phi_D\rangle= \left( \begin{array}{c} 0\\
0 \end{array} \right).\label{vacu}
\end{equation}
With these choices  the model  possesses an exact  $D$-symmetry, which leads to a conserved quantum number ($D$ parity). Because of that the lightest $D$-odd particle is stable, and constitutes a good candidate for a DM particle.

The particle spectrum of the IDM consists of the Higgs boson $h$ which follows from the $\phi_S$ doublet, and the dark scalars $H$, $A$ and $H^{\pm}$ coming from $\phi_D$. The Higgs boson has all tree-level couplings to fermions and gauge bosons equal to the SM ones. Nonetheless, some non-SM effects can occur at the loop level, due to the existence of new scalars. The dark scalars do not couple to fermions at tree level but they do interact with the gauge bosons (through the covariant derivative) and the Higgs particle. The lightest one among them that is neutral plays the role of the DM particle. Here we assume that $M_H<M_A, \m$, hence $H$ is the DM candidate in our model.

Deviations from the SM properties of the Higgs boson can be observed in two ways, because of decays of the Higgs into invisible dark particles or because of additional loop effects thereof. In the following we will first discuss invisible decays of the Higgs boson, and then loop induced decay of the Higgs boson to a pair of photons.

\subsection{Invisible Higgs decays}

The Higgs boson of the IDM has additional, non-SM decay channels leading to dark particles: $h\to 
AA, HH {\rm{\,\,or}}\,\, H^{\pm}H^{\mp}$. The last channel is excluded (at tree-level) by the LEP limits for $\m$: $\m\gtrsim 70\g$. The partial decay width for the process $h\to HH$ reads (see e.g. Ref.~\cite{jhep})
\be\label{part}
\Gamma(h\to HH)=\frac{\lambda_{345}^2 v^2}{32\pi M_h}\sqrt{1-\frac{4M_{H}^2}{M_h^2}},
\ee
where $\lambda_{345}=\lambda_3+\lambda_4+\lambda_5$ is proportional to the coupling between the Higgs boson and a pair of DM particles.  For the decay $h\to AA$ the parameters $\lambda_{345}$ and $M_H$ have to be replaced by $\lambda_{345}^-=\lambda_3+\lambda_4-\lambda_5$ and $M_A$, respectively. 

\begin{figure}[ht]
\centering
\includegraphics[width=.6\textwidth]{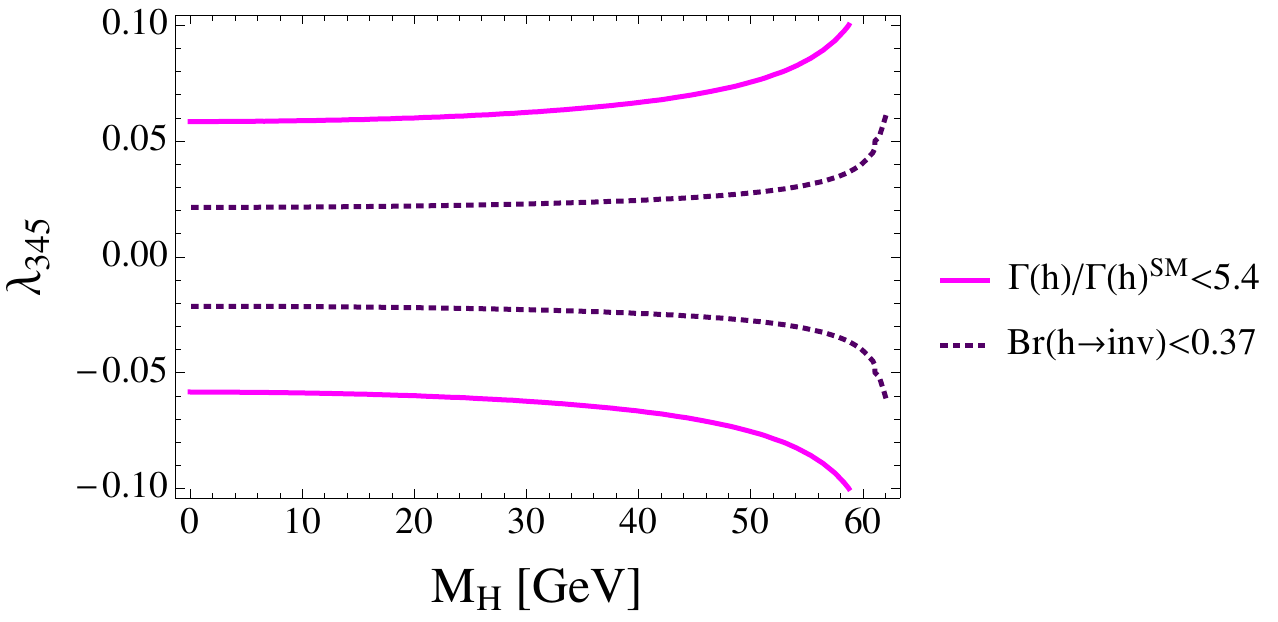}
\caption{Constraints on $\lambda_{345}$ and $M_H$ following from the LHC measurement of Br$(h\to \mathrm{inv})$ and $\Gamma(h)$. The region between the corresponding curves is allowed. We assume that the decay $h\to AA$ is kinematically forbidden.
\label{invisible}}
\end{figure}

Since the decay width~(\ref{part}) depends on the mass of the product of the decay and its coupling to the Higgs boson, these quantities can be constrained with the use of the LHC results on the branching ratio of the Higgs boson decay to invisible particles.
In the same way the measurement of the total Higgs decay width can be used, since $\Gamma (h\to \mathrm{inv})$ contributes significantly to it (see next section). 
 Below, for the sake of simplicity we will assume that $A$ is too heavy for the $h\to AA$ process to be allowed, i.e., $M_H <M_h/2$ and $M_A>M_h/2$. In Fig.~\ref{invisible} the constraints on $\lambda_{345}$ and $M_H$, coming from experimental constraints on Br$(h\to \mathrm{inv})<0.37$~\cite{invisible} and on the total width $\Gamma(h)<5.4\   \Gamma(h)^{\mathrm{SM}}$~\cite{gamma}, are presented. From Fig.~\ref{invisible} one can see that the coupling $\lambda_{345}$ is constrained by Br$(h\to\mathrm{inv})$ to a small value, $|\lambda_{345}|\lesssim0.05$ for $M_H<62\g$.

\subsection{Higgs decays to $\gamma\gamma$}
The differences between the SM and the IDM can also be  observed in the loop induced decays of the Higgs boson, $h\to \gamma\gamma$ and $h\to Z\gamma$. The first of these decays, being measured very precisely,   recently gained much attention, since the first measurements  showed some deviation from the SM expectation giving a hint on the existence of new physics. Nowadays, these measurements converged to the SM, as the observed signal strengths (often denoted by $\mu_{\gamma \gamma}$) are $\rg = 1.17\pm 0.27$ (ATLAS)~\cite{diphoton-ATLAS},  $1.14^{+0.26}_{-0.23}$ (CMS)~\cite{diphoton-CMS}, where the expectation of the SM is $\rg=1$.  We see, than  new physics effects are still acceptable within the experimental bounds. Note that the $Z\gamma$ signal strength has not yet been measured with enough precision to constrain new physics.

Let us consider $R_{\gamma\gamma}$ for the 125 GeV-$h$ in the IDM (see e.g. Refs.~\cite{ rgg, Posch, Arhrib})
\be
R_{\gamma \gamma}:=\frac{\sigma(pp\to h\to \gamma\gamma)^{\textrm{IDM}}}{\sigma(pp\to h\to \gamma\gamma)^{\textrm  {SM}}}\approx\frac{\textrm{Br}(h\to\gamma\gamma)^{\textrm {IDM}}}{\textrm{Br}(h\to\gamma\gamma)^{\textrm {SM}}},
\ee
where we have used the  narrow-width approximation  and the fact that the main production cross section $gg\to h$  is in the IDM the same as in the SM.

\begin{figure}[ht]
\centering
\includegraphics[width=.35\textwidth]{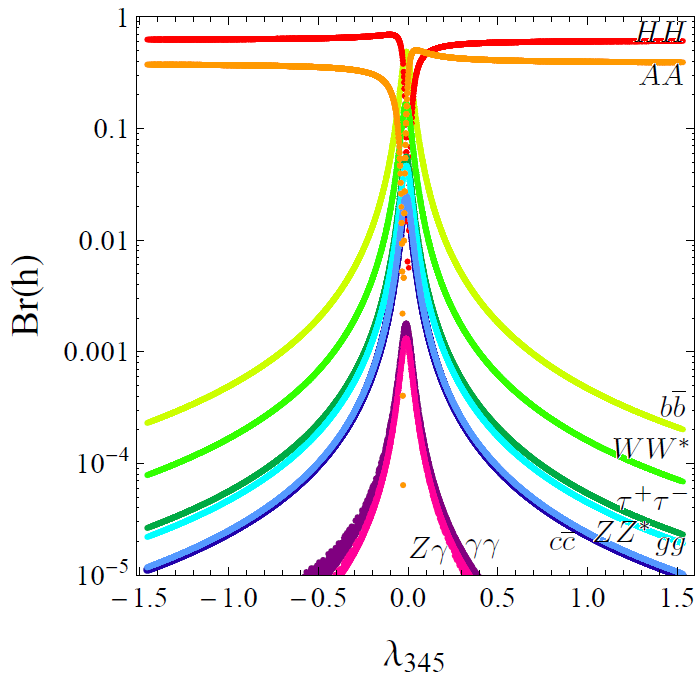}\hspace{.5cm}
\includegraphics[width=.35\textwidth]{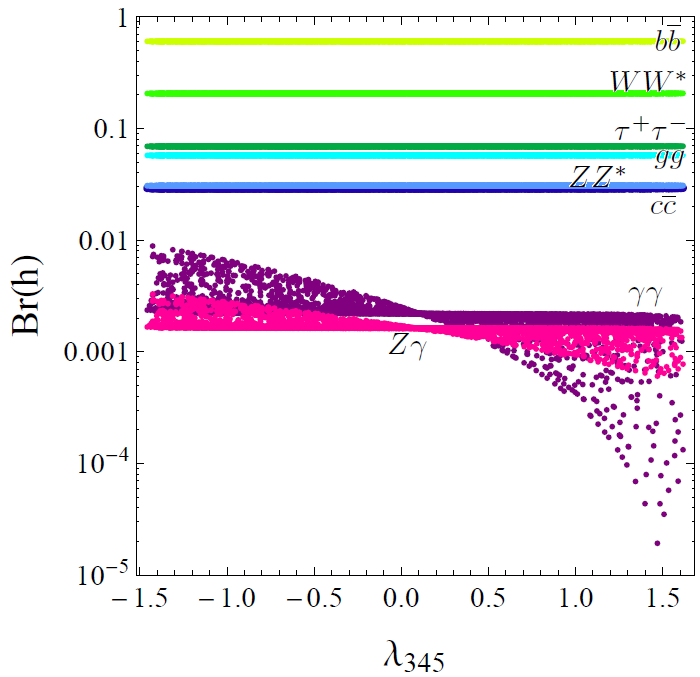}
\caption{Branching ratios of the Higgs boson in the IDM as functions of $\lczp$. Left: invisible channels open ($M_H=50\g$, $M_A=58\g$). Right: invisible channels  closed ($M_H=75\g$, $M_A>M_H$). Figure from Ref.~\cite{rgg}.
\label{br}}
\end{figure}

In the formula above $\textrm{Br}(h\to\gamma\gamma)^{\textrm {SM}}$ is known, and $\textrm{Br}(h\to\gamma\gamma)^{\textrm {IDM}}={\Gamma(h\to\gamma\gamma)^{\textrm {IDM}}}/{\Gamma(h)^{\textrm {IDM}}}$. All the tree-level decay widths of the Higgs boson to SM particles are in the  IDM the same as in the SM. Only the existence of the invisible decay channels, and the $\gamma\gamma$, and $Z\gamma$ decays can modify the total decay width. However,  branching ratios of the latter are very small, at the order of $10^{-3}-10^{-2}$ so they can be ignored, and to a good approximation, only invisible channels modify $\Gamma(h)$ (we used this fact already in Sec.~2.1). The branching ratios in the IDM are presented in Fig.~\ref{br} as functions of $\lambda_{345}$. Note that once the invisible channels are kinematically allowed, they dominate over the SM channels, so in general they tend to suppress $\rg$.

 If invisible channels are closed, the the partial decay width  $\Gamma(h\to \gamma\gamma)$ although small can be a valuable source of information. In the SM the $h\to \gamma\gamma$ decay is induced by the  a $W^{\pm}$ boson loop and  fermionic loops (the top quark dominates).
In general, in the IDM $\Gamma(h\to \gamma\gamma)$  differs from the  SM one because of 
an extra contribution due to  the charged scalar, $H^{\pm}$. This contribution can interfere either constructively or destructively with the SM part. Already in Fig.~\ref{br} (right panel) it is visible that Br$(h\to \gamma\gamma)$ can be enhanced or suppressed with respect to the SM.

\subsubsection{Enhanced diphoton signal strength}

\begin{figure}[ht]
\centering
\includegraphics[width=.45\textwidth]{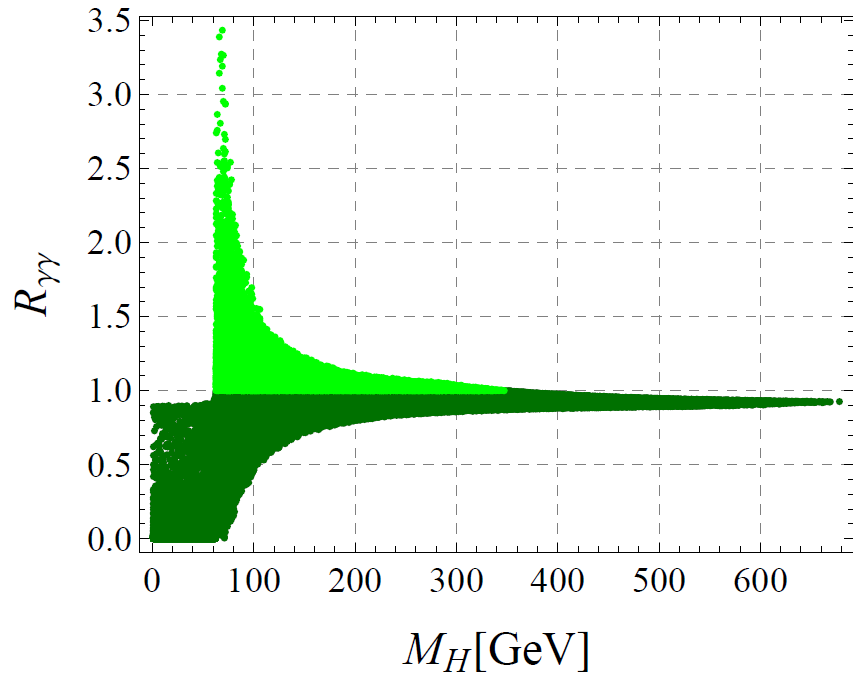}\hspace{.5cm}
\includegraphics[width=.45\textwidth]{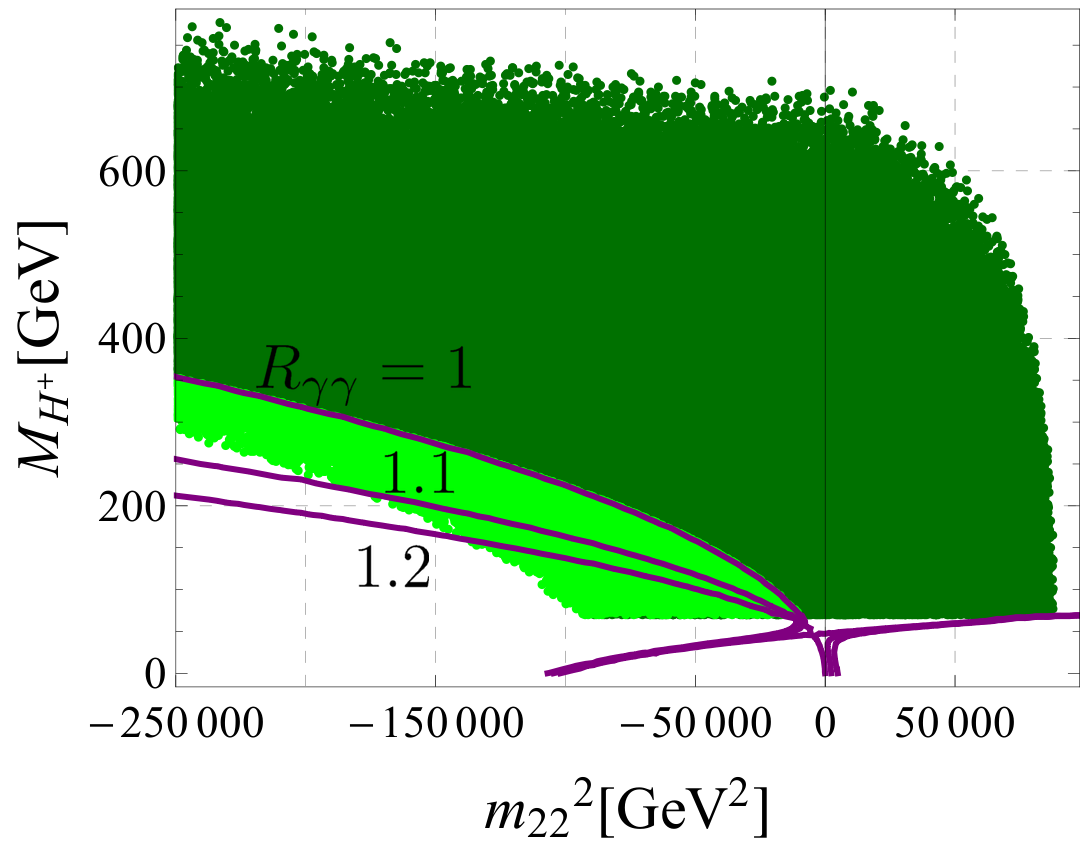}
\caption{Left:  $\rg$ dependence on $M_H$. Right: region allowed by the experimental and theoretical constraints in the $(m_{22}^2,\ \m)$ plane. Light green (gray) indicates the region where $\rg\geqslant1$, the lines correspond to the constant values of $\rg$. Plots are made for $-25\cdot10^4 \g^2\leqslant m_{22}^2\leqslant 9\cdot 10^4\g^2$. From Ref.~\cite{rgg}
\label{rgg-gtr-1}}
\end{figure}

Let us first analyse the consequences of enhanced signal strength (we follow Ref.~\cite{rgg}). In the left panel of Fig.~\ref{rgg-gtr-1}  the dependence of $\rg$ on $M_H$ is shown. One can clearly see that for $M_H<M_h/2\approx 62.5\g$ the diphoton signal strength is always suppressed with respect to the SM. This means that if enhancement of $R_{\gamma \gamma}$ is observed, DM with mass below 62.5 GeV is excluded.

 In the right panel of Fig.~\ref{rgg-gtr-1} the allowed $(m_{22}^2, \m)$  region,  obtained by scanning the parameter space subject to relevant theoretical and experimental constraints\footnote{Such as perturbative unitarity, stability of the Inert vacuum as well as the LEP limits and the EW precision data ($S,T$ parameters).},  is presented. The parameter $m_{22}^2$ is important for $\rg$
  because the coupling between the Higgs boson and the charged scalar is proportional to $2\m^2+m_{22}^2$. In the region marked by light green (gray) $\rg>1$, while the (purple) lines indicate constant values of $\rg$. Note that for  $\rg\geq1$ the viable region is unconstrained, however for substantial enhancement of $\rg$ the allowed region is bounded. For example for $\rg>1.2$,  only fairly light charged scalar (and since $M_H<\m$ also DM) is allowed, $\m, M_H \lesssim 154\g$. The case where $\rg$ goes below 1 will be analysed in the next section, and combined with the DM astrophysical measurements.
\subsection{DM constraints from the Higgs LHC and Planck data }
The current Planck $3\sigma$ limit for DM relic density is $0.1118<\Omega_{DM} h^2<0.1280$~\cite{dm-relic}. $\Omega_{DM} h^2$ depends on DM annihilation and production channels, so this measurement constrains the mass and couplings of the DM candidate.
The IDM is a so-called ``Higgs-portal'' DM model, i.e., in a wide range of masses the DM candidate couples to fermions mainly through the exchange of $h$. Therefore the coupling $\lczp$ between the Higgs and the DM candidate is constrained by relic density measurement. On the other hand the same coupling, as was shown before, is important for the diphoton signal strength. This gives us an opportunity to combine these two types of constraints. In the following we will examine the case $\rg>0.7$ (with agreement with $3 \sigma$ LHC limit), we studied other cases  in~\cite{jhep}.

\begin{figure}[ht]
\centering
\includegraphics[width=.45\textwidth]{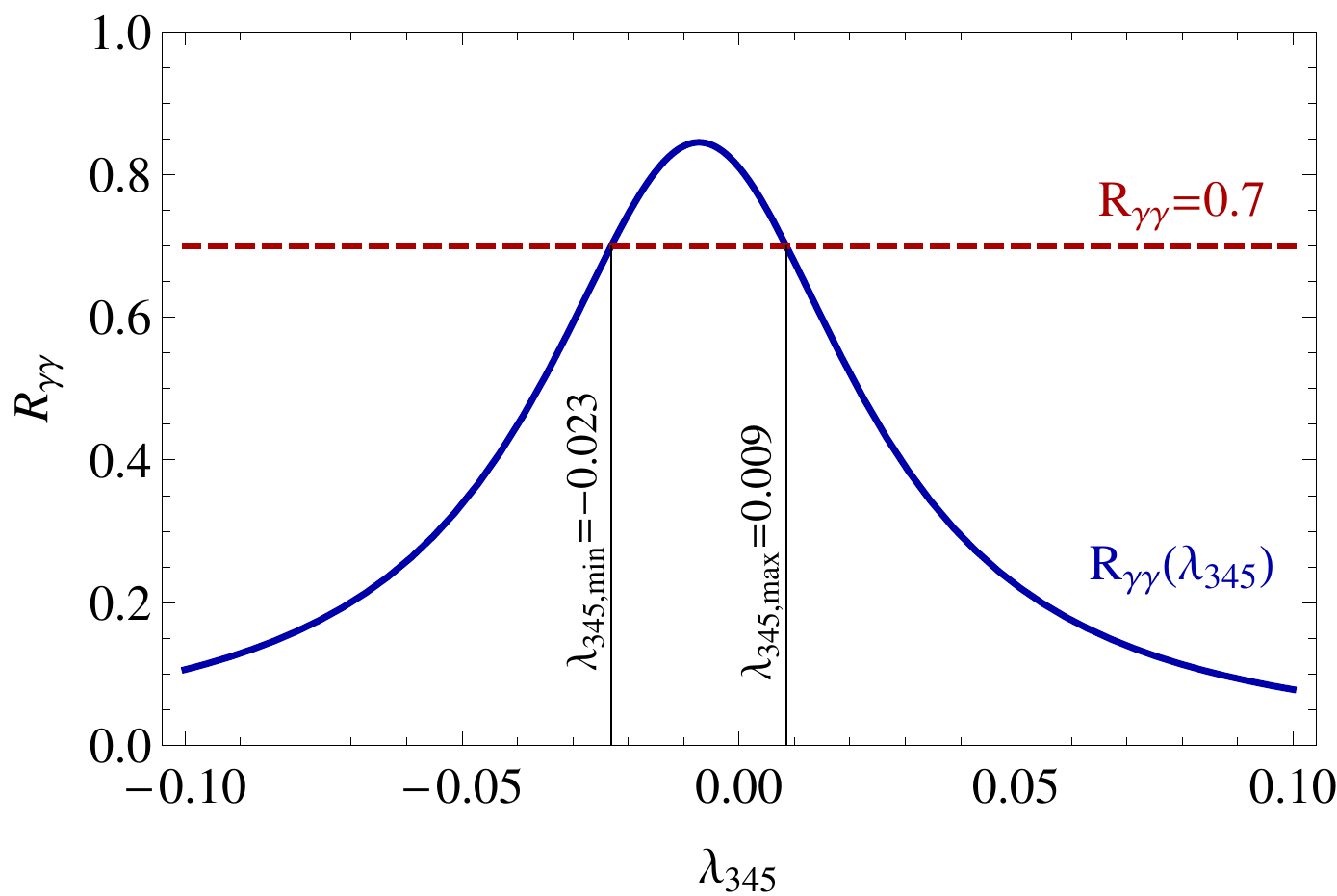}\hspace{.5cm}
\includegraphics[width=.45\textwidth]{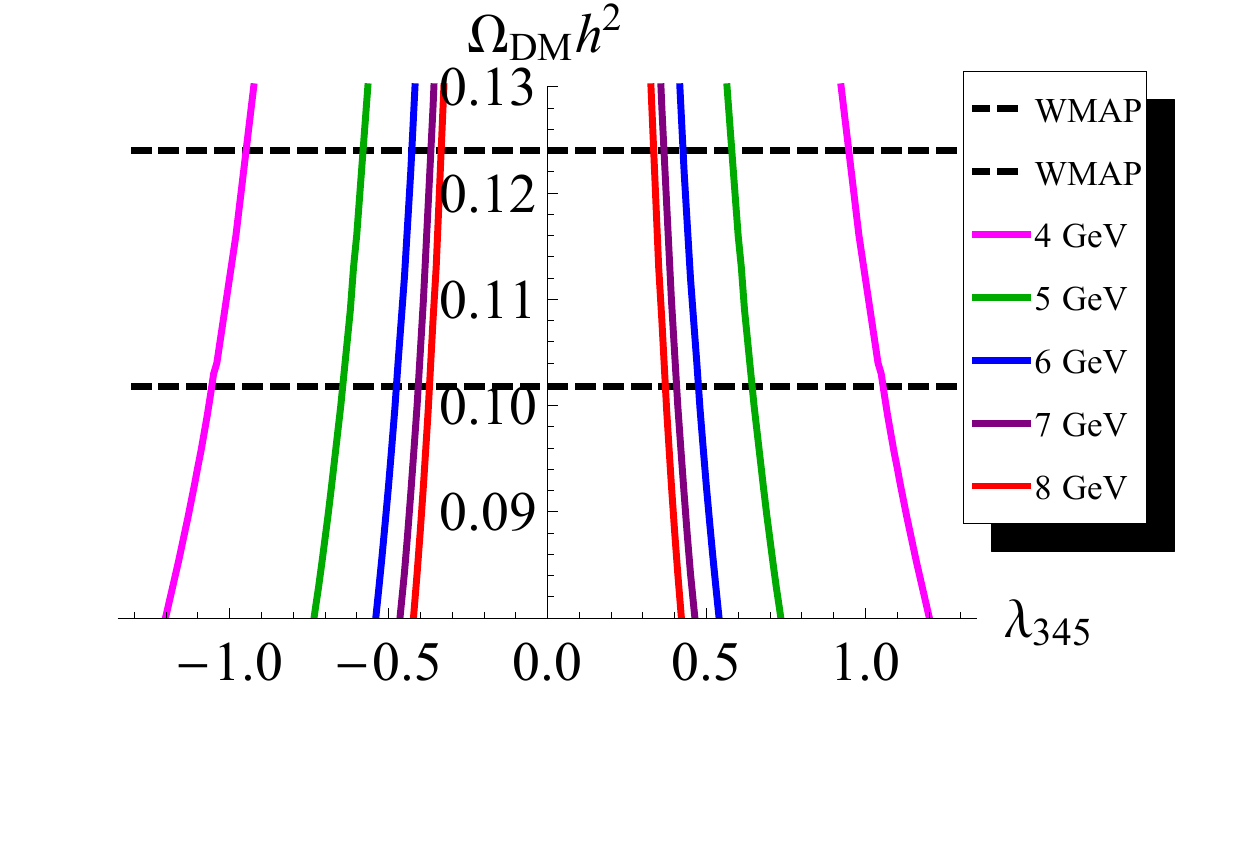}
\caption{Left: $\rg$ as a function of $\lczp$ for $M_H=55\g, M_A=60\g, \m=120\g$ (from  Ref.~\cite{jhep}). Right: relic density of DM as a function of $\lczp$ for different  DM mass. The WMAP 3-$\sigma$ bound is marked by the dashed black horizontal lines (from Ref.~\cite{dm}).
\label{l345-constraints}}
\end{figure}

Fig.~\ref{l345-constraints} shows how the constraints arise. In the left panel $\rg$ as a function of $\lczp$ is shown (for fixed values of masses). If we require that $\rg>0.7$, upper and lower bounds on $\lczp$ arise.
In the right panel the relic density of the DM as a function of $\lczp$ is presented for different values of $M_H$. To fall within the $\Omega_{DM}h^2$ experimental limits (i.e., between the black dashed horizontal  lines)\footnote{On this illustrative plot the WMAP limits are presented but later on we will use the more accurate Planck results.} the value of $\lczp$  should be between the upper and lower limits. These two types of bounds will be combined in the following.

It has been shown in previous works~\cite{dm-inne, dm} that DM in the IDM can have the  correct relic abundance only in three regions: for very light DM ($M_H\lesssim 10\g$), intermediate DM ($40\g \lesssim M_H \lesssim 160 \g$), and heavy DM ($M_H\gtrsim 500\g$). We will analyse these cases separately, following~Ref.~\cite{jhep}.

As can be seen in the right panel of Fig.~\ref{l345-constraints} the right $\Omega_{DM}h^2$
 of very light DM is obtained for $|\lczp|\sim \mathcal{O}(0.5)$. 
Smaller coupling means that DM does not annihilate efficiently enough, and the relic abundance is too big. As~$\lczp$  in agreement with the LHC limit $\rg>0.7$ is around $|\lczp|<0.04$, those two requirements cannot be reconciled, and the very light DM is excluded.
\begin{figure}[h]
\begin{center}
\includegraphics[width=.32\textwidth]{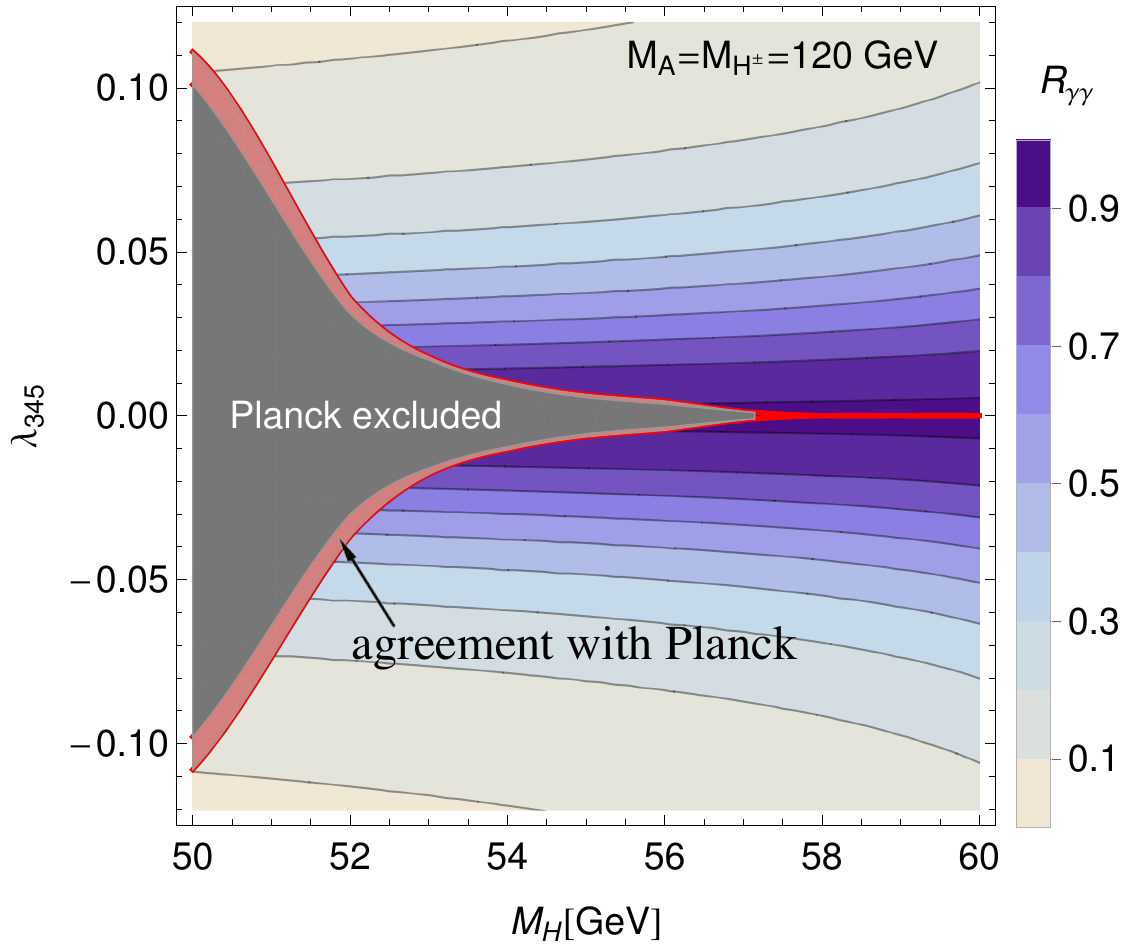}
\includegraphics[width=.32\textwidth]{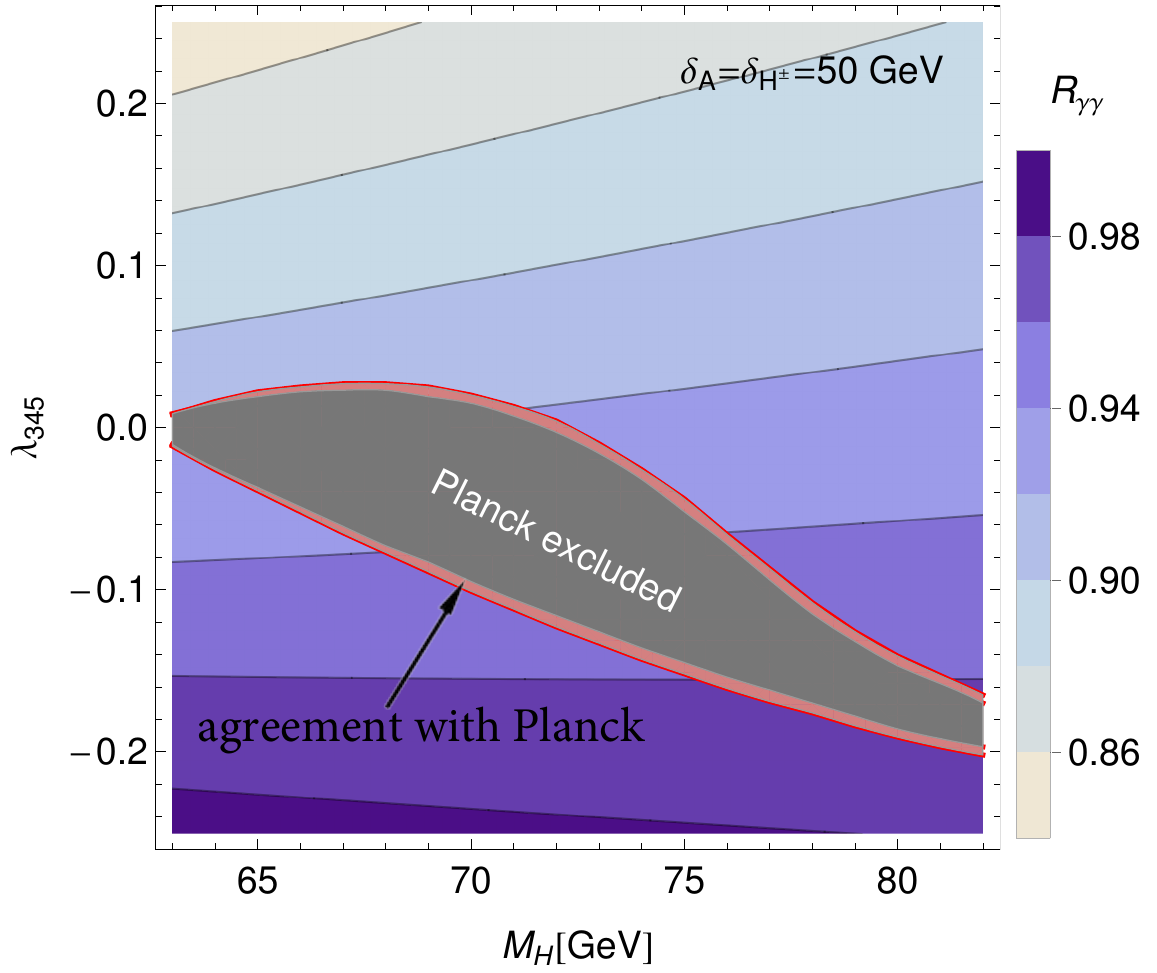}
\includegraphics[width=.32\textwidth]{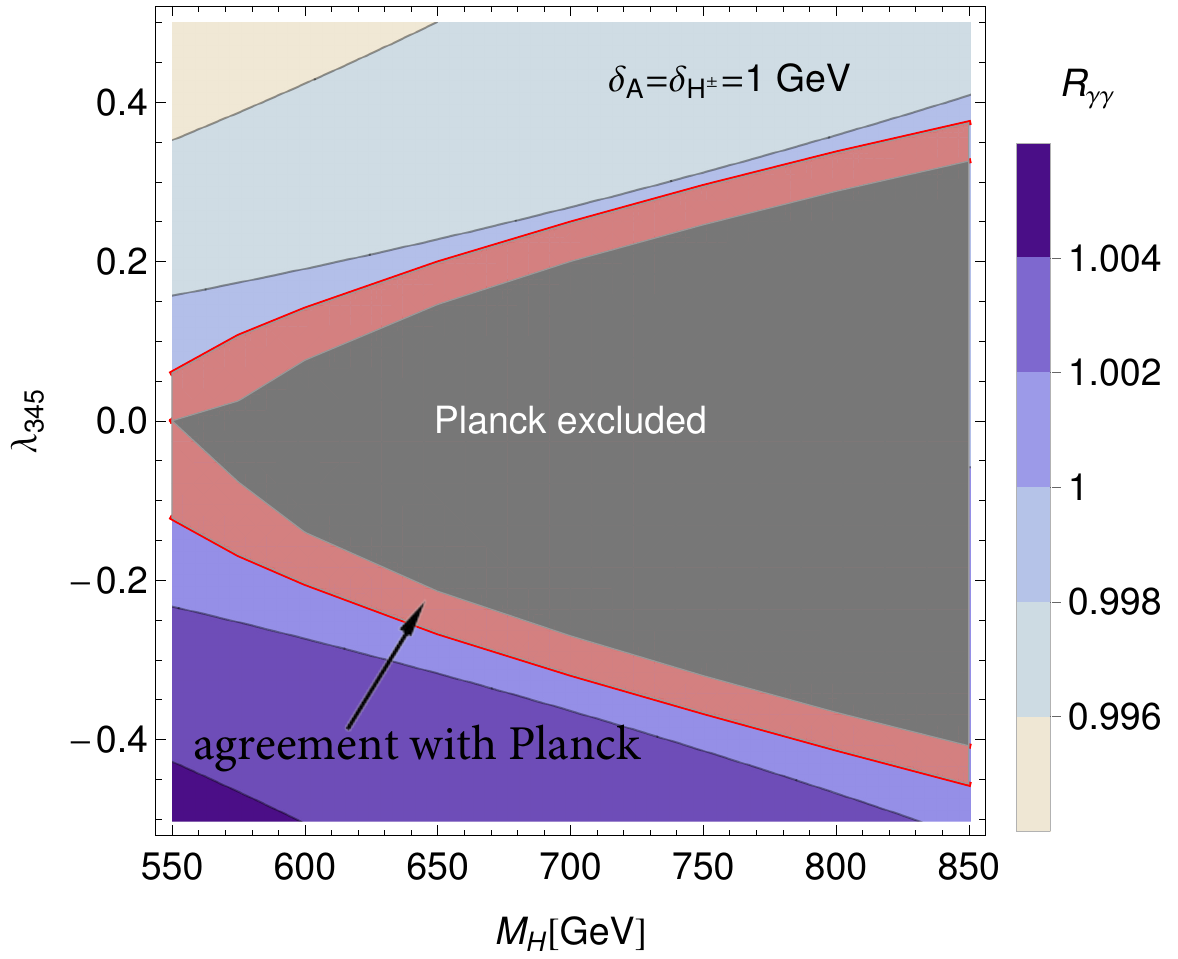}
\end{center}
\caption{Maps of the values of $\rg$ in the $(M_H,\ \lczp)$ plane  for the intermediate DM with $h\to HH$ channel open (left), $h\to HH$ channel closed  (middle), and for heavy DM (right) in comparison with the allowed by Planck (red) bands. See Ref.~\cite{jhep}.\label{fig:dm}}
\end{figure}
Results for intermediate and heavy masses are presented in Fig.~\ref{fig:dm}. The shades (of blue) indicate the values of $\rg$. On this, constraints from Planck are superposed. The dark gray {\underline {inner}} region is excluded ($\Omega_{DM}h^2$ is too big).
 The two regions indicated by arrows/red  bands are  in agreement with Planck data (correct relic density), and in the remaining region the relic density is too low (another DM component  would be necessary in order to comply 
with relic density data).
In the left panel a plot for intermediate DM, with $M_H<M_h/2$ is presented. One can see that relic density constraints (Planck) are in agreement with the assumption $\rg>0.7$ only for $M_H>53\g$. In the middle panel intermediate DM with $M_H>M_h/2$ is analysed. Here all the points that are in agreement with the Planck measurement also give $\rg>0.7$. However, if Planck constraints are to be met, no enhancement in $\rg$ is possible. For the heavy DM (right panel of Fig.~\ref{fig:dm}) we can get correct relic density for all values of masses. Note that $\rg$ is very close to 1 for this case.

\subsection{Comparison with direct DM detection experiments}
\begin{figure}[ht]
\centering
\includegraphics[width=.5\textwidth]{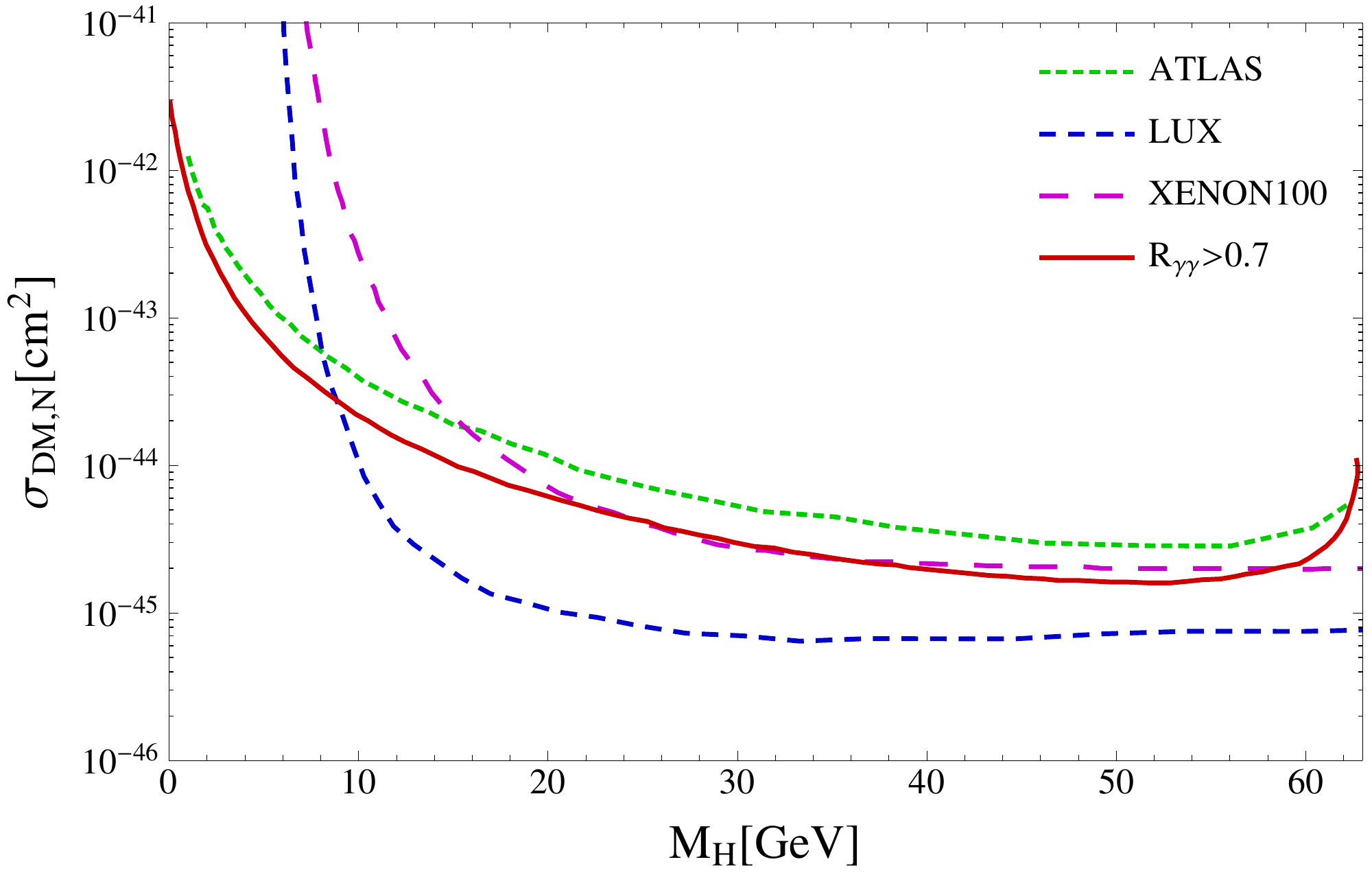}\hspace{.5cm}
\caption{Our results for upper limits on $\sigma_{DM,N}$  coming from the limit $\rg>0.7$ (and $f_n=0.326$) compared with  upper limits from LUX and XENON100 experiments, and from the LHC constraints (ATLAS) on the  Higgs invisible Br. 
\label{lux}}
\end{figure}

The constraints obtained above can be compared with the results of the direct experimental search of the DM~\cite{jhep}, where the DM is supposed to scatter off the nuclei. In the 
Higgs-portal models, among them   
IDM, the cross section $\sigma_{DM,N}$ is proportional to the square of the coupling of Higgs to DM ($\lczp^2$)
because the DM  interacts with the nucleus  through the exchange of the Higgs boson, $\sigma_{DM,N}\sim \lczp^2 f_N^2 /(M_N+M_H)^2 $, where $f_N$ is a formfactor, and $M_N$ is the mass of the nucleon.
 In Fig.~\ref{lux} a~comparison of our results   coming from the limit $\rg>0.7$ , and the constraints from direct DM search experiments (LUX and XENON100), and from constraints on the invisible Higgs branching ratio (LHC  ATLAS)  are presented. Note, that our upper limits, represented by the line $R_{\gamma \gamma}>$ 0.7, are competitive with the upper limits  from the mentioned dedicated DM experiments. 

\begin{figure}[ht]
\centering
\includegraphics[width=.45\textwidth,height=0.195\textheight]{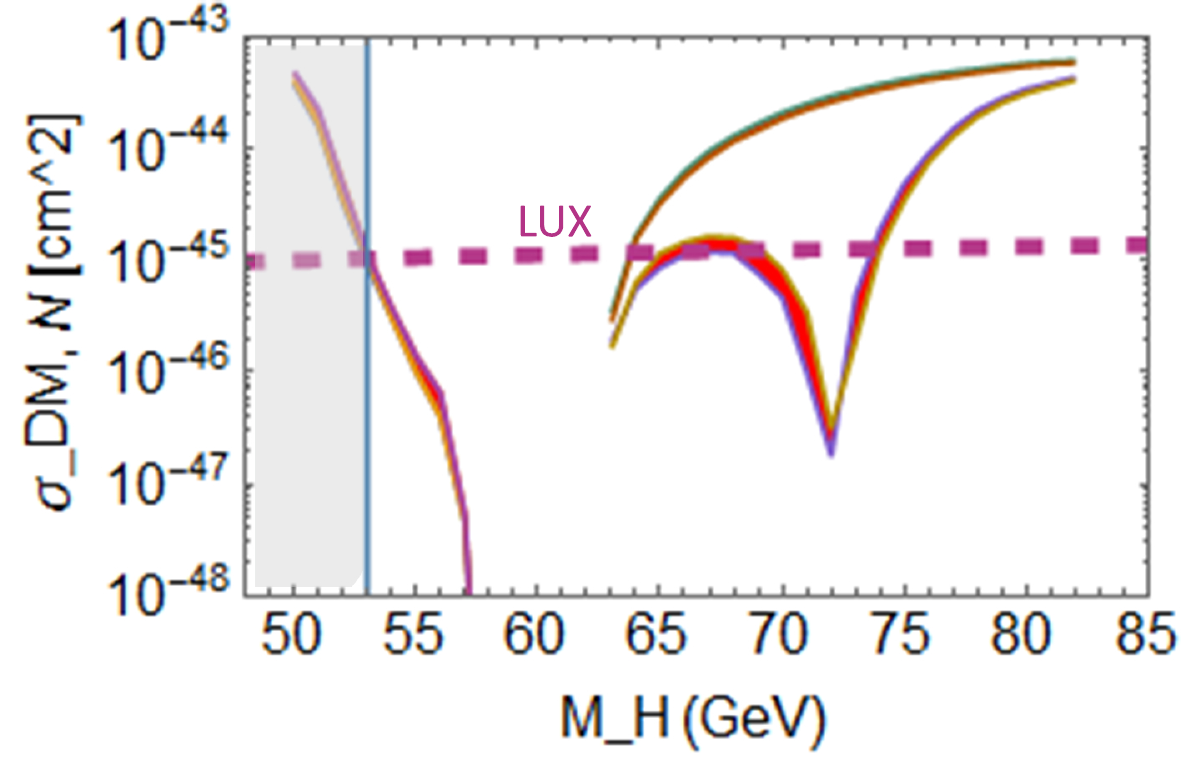}\hspace{.5cm}
\includegraphics[width=.45\textwidth, height=0.2\textheight]{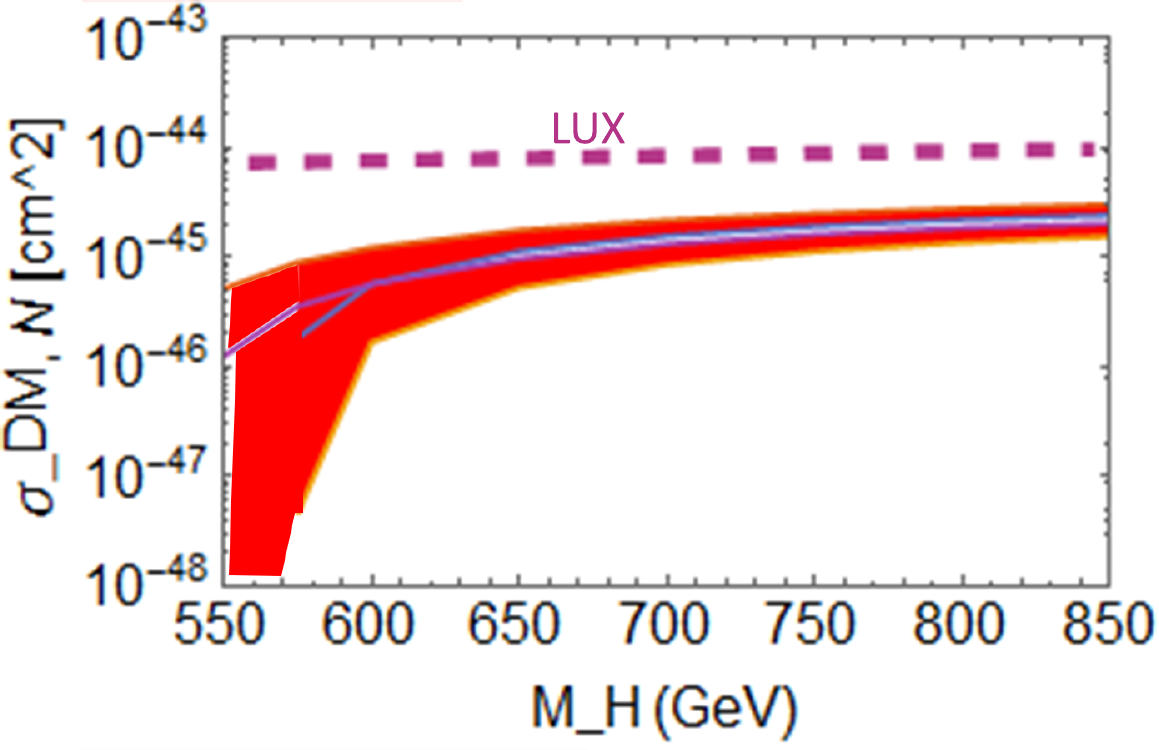}\hspace{.5cm}
\caption{Allowed regions of  $\sigma_{DM,N}$ coming from the Planck limit on the DM (red bands),  compared with upper limits  from LUX (and $f_N=0.326$).  For lower mass part of allowed regions ($M_H < 53$ GeV) is excluded by $\rg>0.7$ condition. 
\label{lux2}}
\end{figure}

Similarly, the constraints coming from the DM relic density measurements (red bands in Fig.~\ref{fig:dm}) can be translated to constraints on the DM-nucleon scattering cross section. In Fig.~\ref{lux2} the allowed regions of $\sigma_{DM,N}$ (red bands) as a function of $M_H$ are shown. They are  coming from $\lambda_{345}$ regions allowed by  the Planck data and $\rg>0.7$ ($f_N=0.326$). Comparison with upper limits  from LUX is shown. We see, that 
the direct detection limits (LUX) stay in agreement with these constraints~\cite{Stal}, however loop corrections can bring the model close to the future experiments reach~\cite{Klasen}.

\section{Conclusions}
The discovery of the Higgs boson  was awaited for a long time since it was the last component needed to complete the Standard Model. Moreover, it also opens door to exploration of new phenomena. The search for new particles at the LHC gives exciting perspectives, but we can also use available data, e.g., the measurements of the Higgs boson properties, especially the  $\gamma\gamma$ signal strength, to shed light on such issues as the properties of the DM.
Other dedicated analysis of the DM in the IDM are ongoing, e.g. on the lepton pair production at the LHC in the processes $q\bar{q}\to HA$ followed by $A\to HZ$ or $H \to l \bar{l}$ ~\cite{s-il}. Finally, models like IDM can shed some light also on the problem of the thermal evolution of the Universe~\cite{ginz}.

\section*{Acknowledgments}
MK would like to thank organizers of this interesting symposium for invitation, warm hospitality and exceptional atmosphere. We thank P.~Swaczyna for his contribution to this work. The work of MK and DS was partially supported by the grant NCN OPUS 2012/05/B/ST2/03306
(2012-2016). The work of B\'{S} was supported by the Polish National Science Centre grant PRELUDIUM, under the decision number DEC-2013/11/N/ST2/04214.

%

\end{document}